\documentclass[aps,prl,reprint,superscriptaddress ]{revtex4-1}

\usepackage{amsmath,amssymb}
\usepackage{wasysym}
\usepackage{graphicx}
\usepackage{dcolumn}
\usepackage{bm}
\usepackage{color}

\begin{document}


\title{Evidence for $Z$=6 `magic number' in neutron-rich carbon isotopes}




\author{D.T.~Tran}
\affiliation{RCNP, Osaka University, Osaka 567-0047, Japan.}
\affiliation{Institute of Physics, Vietnam Academy of Science and Technology, Hanoi 10000, Vietnam.}

\author{H.J.~Ong}
\email{Correspondence: onghjin@rcnp.osaka-u.ac.jp}
\affiliation{RCNP, Osaka University, Osaka 567-0047, Japan.}


\author{G.~Hagen}
\affiliation{Physics Division, Oak Ridge National Laboratory, Oak Ridge, Tennessee 37831, USA.}
\affiliation{Department of Physics and Astronomy, University of Tennessee, Knoxville, Tennessee 37996, USA}

\author{T.~D.~Morris}
\affiliation{Physics Division, Oak Ridge National Laboratory, Oak Ridge, Tennessee 37831, USA.}
\affiliation{Department of Physics and Astronomy, University of Tennessee, Knoxville, Tennessee 37996, USA}

\author{N.~Aoi}
\affiliation{RCNP, Osaka University, Osaka 567-0047, Japan.}

\author{T.~Suzuki}
\affiliation{Department of Physics, College of Humanities and Sciences, Nihon University, Tokyo 156-8550, Japan.}
\affiliation{National Astronomical Observatory of Japan, Tokyo 181-8588, Japan.}

\author{Y.~Kanada-En'yo}
\affiliation{Department of Physics, Kyoto University, Kyoto 606-8502, Japan.}

\author{L.S.~Geng}
\affiliation{School of Physics and Nuclear Energy Engineering, Beihang University, Beijing 100191, China}

\author{S.~Terashima}
\affiliation{School of Physics and Nuclear Energy Engineering, Beihang University, Beijing 100191, China}

\author{I.~Tanihata}
\affiliation{RCNP, Osaka University, Osaka 567-0047, Japan.}
\affiliation{School of Physics and Nuclear Energy Engineering, Beihang University, Beijing 100191, China}

\author{T.T.~Nguyen}
\affiliation{Pham Ngoc Thach University of Medicine, Ho Chi Minh 700000, Vietnam}
\affiliation{Faculty of Physics and Engineering, Ho Chi Minh City University of Science, Ho Chi Minh City 70250, Vietnam.}

\author{Y.~Ayyad}
\affiliation{RCNP, Osaka University, Osaka 567-0047, Japan.}

\author{P.Y.~Chan}
\affiliation{RCNP, Osaka University, Osaka 567-0047, Japan.}

\author{M.~Fukuda}
\affiliation{Department of Physics, Osaka University, Osaka 560-0043, Japan}

\author{H.~Geissel}
\affiliation{GSI Helmholtzzentrum f\"{u}r Schwerionenforschung, D-64291 Darmstadt, Germany.}
\affiliation{Justus Liebig University, 35392 Giessen, Germany.}

\author{M.N.~Harakeh}
\affiliation{KVI Center for Advanced Radiation Technology, University of Groningen, 9747 AA Groningen, The Netherlands.}

\author{T.~Hashimoto}
\affiliation{Rare Isotope Science Project, Institute for Basic Science, Daejeon 34047, Korea.}

\author{T.H.~Hoang}
\affiliation{RCNP, Osaka University, Osaka 567-0047, Japan.}
\affiliation{Institute of Physics, Vietnam Academy of Science and Technology, Hanoi 10000, Vietnam.}

\author{E.~Ideguchi}
\affiliation{RCNP, Osaka University, Osaka 567-0047, Japan.}

\author{A.~Inoue}
\affiliation{RCNP, Osaka University, Osaka 567-0047, Japan.}

\author{G.~R.~Jansen}
\affiliation{National Center for Computational Sciences, Oak Ridge National Laboratory, Oak Ridge, Tennessee 37831, USA}
\affiliation{Physics Division, Oak Ridge National Laboratory, Oak Ridge, Tennessee 37831, USA.}

\author{R.~Kanungo}
\affiliation{Astronomy and Physics Department, Saint Mary's University, Halifax, NS B3H 3C3, Canada.}

\author{T.~Kawabata}
\affiliation{Department of Physics, Kyoto University, Kyoto 606-8502, Japan.}

\author{L.H.~Khiem}
\affiliation{Institute of Physics, Vietnam Academy of Science and Technology, Hanoi 10000, Vietnam.}

\author{W.P.~Lin}
\affiliation{Institute of Modern Physics, Chinese Academy of Sciences, Lanzhou 730000, China.}

\author{K.~Matsuta}
\affiliation{Department of Physics, Osaka University, Osaka 560-0043, Japan}

\author{M.~Mihara}
\affiliation{Department of Physics, Osaka University, Osaka 560-0043, Japan}

\author{S.~Momota}
\affiliation{Kochi University of Technology, Kochi 782-8502, Japan}

\author{D.~Nagae}
\affiliation{RIKEN Nishina Center, Saitama 351-0198, Japan.}

\author{N.D.~Nguyen}
\affiliation{Dong Nai University, Dong Nai 81000, Vietnam.}

\author{D.~Nishimura}
\affiliation{Tokyo University of Science, Chiba 278-8510, Japan}

\author{T.~Otsuka}
\affiliation{Department of Physics, University of Tokyo, Tokyo 113-0033, Japan.}

\author{A.~Ozawa}
\affiliation{Institute of Physics, University of Tsukuba, Ibaraki 305-8571, Japan.}
 
\author{P.P.~Ren}
\affiliation{Institute of Modern Physics, Chinese Academy of Sciences, Lanzhou 730000, China.}

\author{H.~Sakaguchi}
\affiliation{RCNP, Osaka University, Osaka 567-0047, Japan.}

\author{C.~Scheidenberger}
\affiliation{GSI Helmholtzzentrum f\"{u}r Schwerionenforschung, D-64291 Darmstadt, Germany.}
\affiliation{Justus Liebig University, 35392 Giessen, Germany.}


\author{J.~Tanaka}
\affiliation{RCNP, Osaka University, Osaka 567-0047, Japan.}

\author{M.~Takechi}
\affiliation{Department of Physics, Niigata University, Niigata 950-2181, Japan.}

\author{R.~Wada}
\affiliation{Institute of Modern Physics, Chinese Academy of Sciences, Lanzhou 730000, China.}
\affiliation{Cyclotron Institute, Texas A$\&$M University, Texas 77840, USA.}

\author{T.~Yamamoto}
\affiliation{RCNP, Osaka University, Osaka 567-0047, Japan.}


\date{September 4, 2017}

\begin{abstract}
The nuclear shell structure, which originates in the nearly independent motion
of nucleons in an average potential, provides an important guide for our
understanding of nuclear structure and the underlying nuclear forces. Its most
remarkable fingerprint is the existence of the so-called `magic numbers' of
protons and neutrons associated with extra stability. Although the introduction
of a phenomenological spin-orbit (SO) coupling force in 1949 helped explain the
nuclear magic numbers, its origins are still open questions. 
Here, we present experimental evidence for the smallest SO-originated magic
number (subshell closure) at the proton number 6 in $^{13-20}$C obtained from
systematic analysis of point-proton distribution radii, electromagnetic
transition rates and atomic masses of light nuclei. Performing
{\it ab initio} calculations on $^{14,15}$C,
we show that the observed proton distribution radii and subshell closure
can be explained by the state-of-the-art nuclear theory with chiral
nucleon-nucleon and three-nucleon forces, which are rooted in the quantum
chromodynamics.
\end{abstract}


\maketitle


\indent Atomic nuclei -- the finite quantum many-body systems consisting of
protons and neutrons (known collectively as nucleons) -- exhibit shell
structure, in analogy to the electronic shell structure of atoms. Atoms with
filled electron shells -- known as the noble gases -- are particularly stable
chemically. The filling of the nuclear shells, on the other hand, leads to the
magic-number nuclei. The nuclear magic numbers, as we know in stable and
naturally-occurring nuclei, consist of two different series of numbers. The
first series -- 2, 8, 20 -- is attributed to the harmonic oscillator (HO)
potential, while the second one -- 28, 50, 82, and 126 -- is due to the
spin-orbit (SO) coupling force (see Fig. 1). It was the introduction of this SO
force -- a force that depends on the intrinsic spin of a nucleon and its
orbital angular momentum, and the so-called $j$-$j$ coupling scheme that helped
explain~\cite{MayerPR75,JensenPR75} completely the magic numbers, and won
Goeppert-Mayer and Jensen the Nobel Prize. However, the microscopic origins of
the SO coupling force have remained unresolved due to the difficulty to
describe the structure of atomic nuclei from {\it ab initio} nuclear
theories~\cite{BarrettPPNP69,CarlsonRMP87,HagenRPP77} with two- (NN) and
three-nucleon forces (3NFs). Although the theoretical study~\cite{PieperPRL70}
of the SO splitting of the $1p_{1/2}$ and $1p_{3/2}$ single-particle
states in $^{15}$N has suggested possible roles of two-body SO and
tensor forces, as well as three-body forces, the discovery of the so-far elusive
SO-type magic number 6 is expected to offer unprecedented opportunities to
understand its origins.

\indent In her Nobel lecture, Goeppert-Mayer had mentioned the magic numbers
6 and 14 -- which she described as ``hardly noticeable'' -- but
surmised that the energy gap between the $1p_{1/2}$ and $1p_{3/2}$ orbitals due
to the SO force is ``fairly small''~\cite{MayerNobel}. That the $j$-$j$ coupling
scheme appears to fail in the $p$-shell light nuclei was discussed and
attributed to their tendency to form clusters of nucleons~\cite{InglisRMP25}.
Experimental and theoretical studies in recent decades, however, have hinted at
the possible existence of the magic number 6 in some semimagic unstable nuclei,
each of which has a HO-type magic number of the opposite type of nucleons. For
instance, possible subshell closures have been suggested in
$^{8}$He~\cite{OtsukaPRL87,SkazaPRC73}, $^{14}$C and
$^{14}$O~\cite{AngeliADNDT99}. Whether such subshell closures exist, and
are predominently driven by the shell closures of the counterpart
protons/neutrons are of fundamental importance.

\indent The isotopic chain of carbon -- with six protons and consisting of
thirteen particle-bound nuclei, provides an important platform to study the
SO splitting of the $1p_{1/2}$ and $1p_{3/2}$ orbitals. Like other lighter
isotopes, the isotopes of carbon are known to exhibit both clustering and
single-particle behaviours. While the second excited $J^\pi$=$0^+$ state in
$^{12}$C -- the famous Hoyle state and important doorway state that helps
produce $^{12}$C in stars -- is well understood as a triple-alpha state, it
seems that the effect of the alpha-cluster-breaking $1p_{3/2}$ subshell closure
is important to reproduce the ground-state binding energy~\cite{SuharaPRC91}.
For even-even neutron-rich carbon isotopes, theoretical calculations using the
anti-symmetrized molecular dynamics (AMD)~\cite{EnyoPRC91},
shell model~\cite{FujimotoPHD2003,SFujiPLB650}, as well as the {\it ab-initio}
no-core shell model calculation~\cite{ForssenJPG40} with NN+3NFs have
predicted near-constant proton distributions, a widening gap between proton
$1p_{1/2}$ and $1p_{3/2}$ single-particle orbits, and a remarkably low proton
occupancy in the $1p_{1/2}$ orbit, respectively. Experimentally, small
$B$($E2$) values comparable to that of $^{16}$O were reported from the
lifetime measurements of the first excited $2^+$ ($2^+_1$) states in
$^{16,18}$C~\cite{WiedekingPRL100,OngPRC78}. The small $B$($E2$)
values indicate small proton contributions to the transitions, and together
with the theoretical predictions may imply the existence of a proton subshell
closure.

\begin{figure}[htbp]
\includegraphics[scale=0.3]{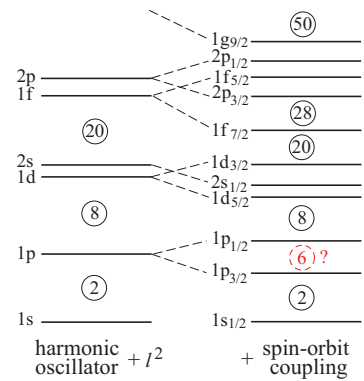}
\caption{\label{fig:shell} Nuclear shell structure for a harmonic oscillator
  potential plus a small orbital angular momentum ($l^2$) term. The right 
  diagram shows the splitting of the single-particle orbitals by an additional
  spin-orbit coupling force.}
\end{figure}

\indent Here we present experimental evidence for a proton-subshell closure
at $Z$=6 in $^{13-20}$C, based on a systematic study of (i) point-proton
distribution radii obtained from our recent experiments as well as the existing
nuclear charge radii~\cite{AngeliADNDT99}, (ii) electric quadrupole transition
rates $B$($E2$) between the $2^+_1$ and ground ($0^+_{\rm gs}$) states of
even-even nuclei~\cite{PritychenkoADNDT107}, and (iii) atomic-mass
data~\cite{WangCPC36}. We show, by performing coupled-cluster calculations, that
the observations are supported by the {\it ab-initio} nuclear model that employs
the nuclear forces derived from the effective field theory of the quantum
chromodynamics.

\indent Although still not well established, the size of a nucleus, which can
be defined as the root-mean-square (rms) radius of its nucleon distribution, is
expected to provide important insights on the evolution of the magic numbers.
Recently, an unexpectedly large proton rms radius (denoted simply as
`proton radius' hereafter) was reported~\cite{GarciaRuizNature12}, and 
suggested as a possible counter-evidence for the double shell closure in
$^{52}$Ca~\cite{WienholtzNature498}. Attempts to identify any emergence of
non-traditional magic numbers based on the analysis of the systematics of the
experimental proton radii have been reported~\cite{AngeliADNDT99,AngeliJPG42}.
For the 4$< Z <$ 10 region, the lack of experimental data on the proton radii of
neutron-rich nuclei due to the experimental and theoretical limitations of the
isotope-shift method has hindered systematic analysis of the radii behaviour.
Such systematic analysis has become possible very recently owing to the
development of an alternative method to extract the proton radii of
neutron-rich nuclei from the charge-changing cross-section measurements.

\indent The charge-changing cross section (denoted as $\sigma_{\rm CC}$) of a
projectile nucleus on a nuclear/proton target is defined as the total cross
section of all processes that change the proton number of the projectile
nucleus. Applying this method, we have determined the proton radii of
$^{14}$Be~\cite{TerashimaPTEP101}, $^{12-17}$B~\cite{EstradePRL113} and
$^{12-19}$C~\cite{KanungoPRL117} from the $\sigma_{\rm CC}$ measurements at GSI,
Darmstadt, using secondary beams at around 900 MeV per nucleon.
In addition, we have also measured $\sigma_{\rm CC}$'s for $^{12-18}$C on a
$^{12}$C target with secondary beams at around 45 MeV per nucleon at the exotic
nuclei (EN) beam line at RCNP, Osaka University. To extract proton radii from
both low- and high-energy data, we have devised a global parameter set for use
in the Glauber model calculations. The Glauber model thus formulated was
applied to the $\sigma_{\rm CC}$ data at both energies to determine the proton
radii. All experimental details, data analysis, and Glauber-model
analysis are given in ref.~\cite{TranPRC94}. 

\begin{table}[th]
\caption{\label{tab:cccs}
  Secondary beam energies and measured $\sigma_{\text{CC}}$'s for $^{17-19}$C.
  The data in the fourth and fifth columns are from ref.~\cite{KanungoPRL117}.
  $R_{\text{p}}$'s in the sixth column are the proton radii
  extracted from the $\sigma_{\text{CC}}$'s in the third and fifth columns. }
\begin{ruledtabular}
\begin{tabular}{cccccc}
  & \multicolumn{1}{c}{$E_{\text{CC}}$} 
  & \multicolumn{1}{c}{$\sigma_{\text{CC}}$}
  & \multicolumn{1}{c}{$E_{\text{CC}}$} 
  & \multicolumn{1}{c}{$\sigma_{\text{CC}}$}
  & \multicolumn{1}{c}{$R_{\text{p}}$}\\
  & \multicolumn{1}{c}{($A$ MeV)}
  & \multicolumn{1}{c}{(mb)}
  & \multicolumn{1}{c}{($A$ MeV)}
  & \multicolumn{1}{c}{(mb)}
  & \multicolumn{1}{c}{(fm)}\\
\hline
$^{17}$C & 46.3 & 1000(16) & 979 & 754(7) & 2.43(4)   \\
$^{18}$C & 42.8 & 1023(31) & 895 & 747(7) & 2.42(5)   \\
$^{19}$C &      &          & 895 & 749(9) & 2.43(4)   \\
\end{tabular}
\end{ruledtabular}
\end{table}

\begin{figure}[htbp]
\includegraphics[scale=0.42]{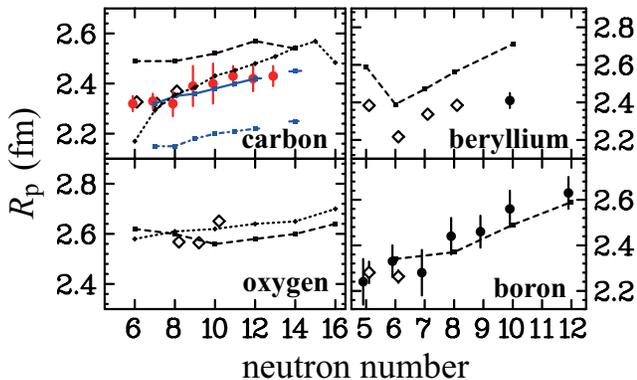}
\caption{\label{fig:rms}
  Proton radii of carbon, beryllium, boron and oxygen isotopes.
  The red- and black-filled circles are, respectively, the proton radii from
  this and our recent
  work~\cite{TerashimaPTEP101,EstradePRL113,KanungoPRL117,TranPRC94}. The open
  diamonds are the data from electron scattering and isotope-shift
  methods~\cite{AngeliADNDT99}. The small symbols connected with dashed and
  dotted lines are the predictions from the AMD~\cite{EnyoPRC91} and
  RMF~\cite{GengPTP113} models, respectively. The small blue symbols with
  solid and dash-dotted lines are the results from the {\it ab-initio}
  coupled-cluster calculations with NNLO$_{\rm sat}$~\cite{EkstromPRC91} and
  the NN-only interaction NNLO$_{\rm opt}$~\cite{EkstromPRL110}.}
\end{figure}

\indent For simplicity, we show only the results for $^{17-19}$C in
Table~\ref{tab:cccs}; for results on $^{12-16}$C, see ref.~\cite{TranPRC94}.
$R_{\text{p}}$'s are the proton radii extracted using the Glauber model
formulated in ref.~\cite{TranPRC94}. The values for $^{17,18}$C
are the weighted mean extracted using $\sigma_{\text{CC}}$'s at the two energies,
while the one for $^{19}$C was extracted using the high-energy data. In
determining the proton radii, we have assumed harmonic-oscillator (HO)-type
distributions for the protons in the Glauber calculations. The uncertainties
shown in the brackets include the statistical uncertainties, the experimental
systematic uncertainties, and the uncertainties attributed to the choice of
functional shapes, i.e. HO or Woods-Saxon, assumed in the calculations.

\indent To get an overview of the isotopic dependence, we compare the proton
radii of the carbon isotopes with those of the neighbouring beryllium, boron
and oxygen isotopes. Figure~\ref{fig:rms} shows the proton
radii for carbon, beryllium, boron and oxygen isotopes. The red- and
black-filled circles are the data for $^{12-19}$C, beryllium and boron isotopes
extracted in this and our previous
work~\cite{TerashimaPTEP101,EstradePRL113,KanungoPRL117,TranPRC94}. For
comparison, the proton radii determined with the electron scattering and
isotope-shift methods~\cite{AngeliADNDT99} are also shown in
Fig.~\ref{fig:rms} (open diamonds). Our
$R_\text{p}$'s for $^{12-14}$C are in good agreement with the electron-scattering
data. In addition, we performed theoretical calculations. The small symbols
connected with dashed and dotted lines shown in the figure are the results from
the AMD~\cite{EnyoPRC91} and relativistic mean field (RMF)~\cite{GengPTP113}
calculations, respectively. The blue-solid and blue-dash-dotted
lines are the results (taken from ref.~\cite{KanungoPRL117}) of
the {\it ab-initio} coupled-cluster (CC) calculations with
NNLO$_{\rm sat}$~\cite{EkstromPRC91} and the NN-only interaction
NNLO$_{\rm opt}$~\cite{EkstromPRL110}, respectively.
The AMD calculations reproduce the trends of all isotopes qualitatively but
overestimate the proton radii for carbon and beryllium isotopes. The RMF
calculations, on the other hand, reproduce most of the proton radii of carbon
and oxygen isotopes but underestimate the one of $^{12}$C. Overall, the CC
calculations with the NNLO$_{\rm sat}$ interactions reproduce the proton radii
for $^{13-18}$C very well. The calculations without 3NFs
underestimate the radii by about 10$\%$, thus suggesting the importance of 3NFs.

\begin{figure}[htbp]
\includegraphics[scale=0.4]{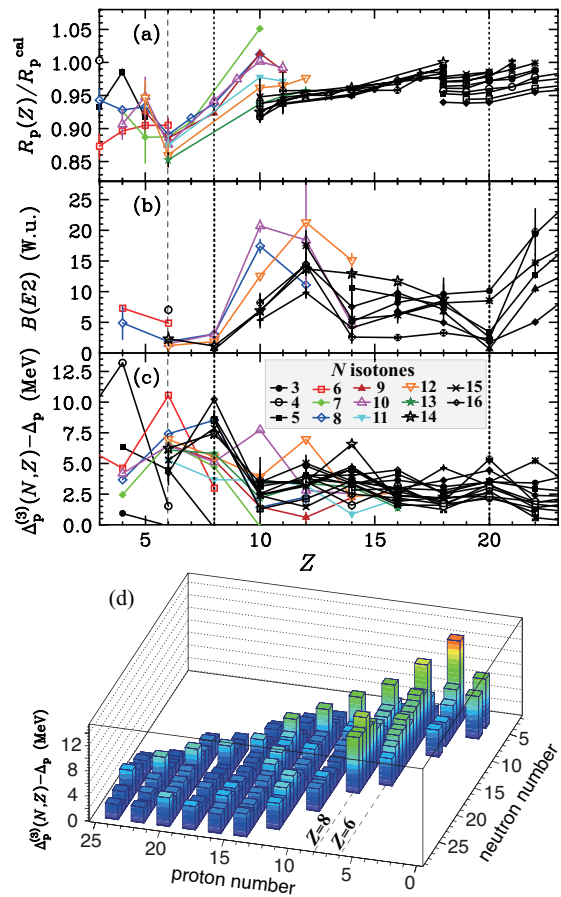}
\caption{\label{fig:systematics} Evolution of
  (a) $R_{\text{p}}/R^{\text{cal}}_{\text{p}}$, (b) $B$($E2$), and
  (c) $\Delta^{(3)}_{\text{p}}(N,Z)-\Delta_{\text p}$ with proton number up to
  $Z$=22 for all available isotonic chains. Vertical dotted lines and
  thin-dashed denote positions of the traditional proton magic numbers and
  $Z$=6, respectively. (d) Two dimensional lego plot of (c).}
\end{figure}

\indent It is interesting to note that $R_\text{p}$'s are almost constant
throughout the isotopic chain from $^{12}$C to $^{19}$C, fluctuating by less
than 5$\%$. Whereas this trend is similar to the one observed/predicted in the
proton-closed shell oxygen isotopes, it is in contrast to those in the
beryllium and boron isotopic chains, where the proton radii change by as much
as 10$\%$ (for berylliums) or more (for borons). It is also worth noting that
most theoretical calculations shown predict almost constant proton radii in
carbon and oxygen isotopes. The large fluctuations observed in Be and B
isotopes can be attributed to the development of cluster structures, whereas
the almost constant $R_{\text{p}}$'s for $^{12-19}$C observed in the
present work may indicate an inert proton core, i.e. $1p_{3/2}$ proton-subshell
closure.

\indent Examining the $Z$ dependence of the proton rms radii along the $N$=8
isotonic chain, Angeli {\it et al.} have pointed
out~\cite{AngeliADNDT99,AngeliJPG42} a characteristic change of slope
(existence of a `kink'), a feature closely associated with shell closure, at
$Z$=6. Here, by combining our data with the recent
data~\cite{TerashimaPTEP101,EstradePRL113,KanungoPRL117,TranPRC94}, as well as
the data from ref.~\cite{AngeliADNDT99}, we plot the experimental
$R_{\text{p}}$'s against proton number. To eliminate the smooth mass number
dependence of the proton rms radii, we normalized all $R_{\text{p}}$'s by the
following mass-dependent rms radii~\cite{CollardElton}:
\begin{equation*}
  R^{\text{cal}}_{\text{p}} = \sqrt{3/5}\left( 1.15 + 1.80A^{-2/3} - 1.20A^{-4/3} \right) A^{1/3} {\rm fm}.
\label{eq:radii}
\end{equation*}  
Figure~\ref{fig:systematics}(a) shows the evolution of
$R_{\text{p}}/R^{\text{cal}}_{\text{p}}$ with proton number up to $Z$=22 for all
available isotonic chains. Each isotonic chain is connected by a solid line.
For clarity, the data for $N$=6 -- 13 isotones are displayed in colours.
For nuclides with more than one experimental value, we have adopted the
weighted mean values. The discontinuities observed at $Z$=10 and $Z$=18 are due
to the lack of experimental data in the proton-rich region. Note the
increase/change in the slope at the traditional magic numbers $Z$=8 and 20.
Marked `kinks', similar to those observed at $Z$=20, 28, 50 and
82~\cite{AngeliJPG42}, are observed at $Z$=6 for isotonic chains from $N$=7 to
$N$=13, indicating a possible major structural change, e.g., emergence of a
subshell closure, at $Z$=6.

\indent The possible emergence of a proton subshell closure at $Z$=6 in
neutron-rich even-even carbon isotopes is also supported by the small $B$($E2$)
values observed in
$^{14-20}$C~\cite{PritychenkoADNDT107,OngPRC78,WiedekingPRL100}.
Figure~\ref{fig:systematics}(b) shows the systematics of $B$($E2$) values
in Weisskopf unit (W.u.) for even-even nuclei up to $Z$=22. All data are
available in ref.~\cite{PritychenkoADNDT107}. 
Nuclei with shell closures manifest themselves as minima. Besides the
traditional magic number $Z$=8, clear minima with $B$($E2$) values smaller
than 3 W.u. are observed at $Z$=6 for $N$=8,10,12 and 14 isotones. 

\indent To further examine the possible subshell closure at $Z$=6, we consider
the second derivative of binding energies defined as follows:
\begin{equation}
  \Delta^{(3)}_{\text{p}}(N,Z) \equiv (-1)^Z [S_{\text{p}}(N,Z) -
    S_{\text{p}}(N,Z+1)],
\label{eq:gap}
\end{equation}
where $S_{\text{p}}(N,Z)$ is the one-proton separation energy. In the absence of
many-body correlations such as pairings, $S_{\text{p}}(N,Z)$ resembles the
single-particle energy, and $\Delta^{(3)}_{\text{p}}(N,Z)$ yields the proton
single-particle energy-level spacing or shell gap between the last
occupied and first unoccupied proton orbitals in the nucleus with $Z$
protons (and $N$ neutrons). To eliminate the effect of proton-proton (p-p)
pairing, we subtract out the p-p pairing energies using the empirical formula:
$\Delta_{\text p} = 12A^{-1/2}$ MeV. Figure~\ref{fig:systematics}(c) shows the
systematics of $\Delta^{(3)}_{\text{p}}(N,Z)-\Delta_{\text p}$ for even-$Z$
nuclides. All data were evaluated from the binding energies~\cite{WangCPC36}.
Here, we have omitted odd-$Z$ nuclides to avoid odd-even staggering effects.
The cusps observed at $Z$=$N$ for all isotonic chains are due to the Wigner
effect~\cite{WignerRPP8}. Apart from the $Z$=$N$ nuclides, sizable `gaps'
($>$ 5 MeV) are also observed at $Z$=6 for $N$=7 -- 15, and at $Z$=8 for
$N$=8 -- 10 and 12 -- 16, although the `gap' appears to decrease from $^{18}$C
to $^{22}$C. For clarity, we show the corresponding two-dimensional lego plot
in Fig.~\ref{fig:systematics}(d).

\indent By requiring a `magic' nucleus to fulfil at least two signatures
in Fig.~\ref{fig:systematics}(a)--(c), we conclude that we have observed a
prominent proton subshell closure at $Z$=6 in $^{13-20}$C. Although
the empirical $\Delta^{(3)}_{\text{p}}$ for $^{12}$C is large ($\sim$14 MeV),
applying the prescription from ref.~\cite{ChasmanPRL99}, we obtain about
10.7 MeV for the total p-p and p-n pairing energy. This estimated large pairing
energy indicates possible significant many-body correlations such as cluster
correlations, which compete and coexist with the single-particle 1$p_{3/2}$
subshell closure in $^{12}$C. Based on this fact, as well as its non-minimum
proton radius and relatively large $B$($E2$) value, we have excluded $^{12}$C
from our conclusion.

\begin{figure}[htbp]
\includegraphics[scale=0.4]{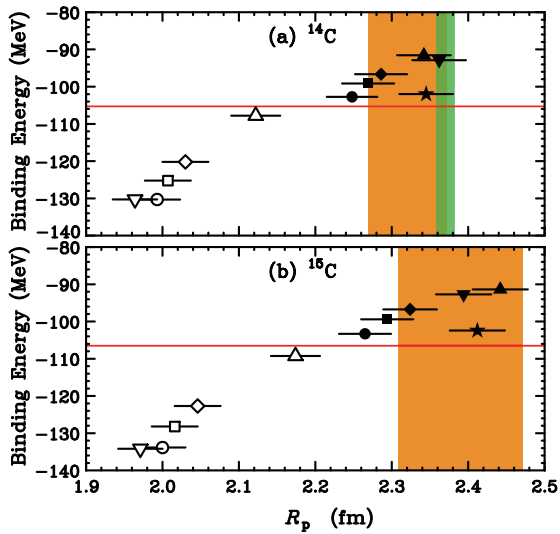}
\caption{\label{fig:sgrms} Binding energies of (a) $^{14}$C and (b) $^{15}$C
  as functions of their proton radii. The coloured bands
  and red-horizontal lines are the experimental values. The green band 
  represents the proton radius from the electron scattering. The filled black
  symbols are the CC predictions with SRG-evolved NN+3NF chiral effective
  interactions at different NN/3NF cutoffs and NNLO$_{\rm sat}$, whereas the
  open symbols are the predictions with the NN-only EM and NNLO$_{\rm sat}$ 
  interactions. See text for details.}
\end{figure}

\indent It is surprising that the systematics of the proton radii, $B$($E2$)
values and the empirical proton `subshell gaps' for most of the carbon isotopes
are comparable to those for proton-closed shell oxygen isotopes. To understand
the observed ground-state properties, i.e. the proton radii and `subshell gap'
of the carbon isotopes, we performed {\it ab-initio} CC calculations on
$^{14,15}$C using various state-of-the-art chiral interactions. We employed the
CC method in the singles-and-doubles approximation with perturbative
triples corrections [$\Lambda$-CCSD(T)]~\cite{TaubeJCP128}
to compute the ground-state binding energies and proton radii for the
closed-(sub)shell $^{14}$C. To compute $^{15}$C ($1/2^{+}$), we used the
particle-attached equation-of-motion CC (EOM-CC) method~\cite{GourPRC74}, and
included up to three-particle-two-hole (3p-2h) and two-particle-three-hole
(2p-3h) corrections. Figure~\ref{fig:sgrms} shows the binding energies as 
functions of the proton radii for (a) $^{14}$C and (b) $^{15}$C. The
coloured bands are the experimental values; the binding energies (red
horizontal lines) are taken from ref.~\cite{WangCPC36}, while proton radii are
from ref.~\cite{TranPRC94} (orange bands) and the electron scattering
data~\cite{AngeliADNDT99} (green band). The filled black symbols are CC
predictions with the NN$+$3NF chiral interactions from ref.~\cite{HebelerPRC83}
labeled 2.0/2.0 (EM)($\blacksquare$),
2.0/2.0 (PWA)($\blacktriangledown$), 1.8/2.0 (EM)($\CIRCLE$),
2.2/2.0 (EM)($\blacklozenge$), 2.8/2.0 (EM)($\blacktriangle$), and
NNLO$_{\rm sat}$~\cite{EkstromPRC91} ($\bigstar$). Here, the NN interactions are
the next-to-next-to-next-to leading-order (N$^3$LO) chiral interaction from
ref.~\cite{EMPRC68}, evolved to lower cutoffs (1.8/2.0/2.2/2.8 fm$^{-1}$)
via the similarity-renormalization-group (SRG) method~\cite{BognerPRC75}, while
the 3NF is taken at NNLO with a cutoff of 2.0 fm$^{-1}$ and adjusted to the
triton binding energy and $^4$He charge radius. The error bars are the
estimated theoretical uncertainties due to truncations of the employed method
and model space~\cite{HagenNatPhys12}. Note that the error bars for the binding
energies are smaller than the symbols. Depending on the NN cutoff, the
calculated binding energy correlates strongly with the calculated proton
radius. In addition, we performed the CC calculations with chiral effective
interactions without 3NFs, i.e. the NN-only EM interactions with NN cutoffs at
1.8 ($\ocircle$), 2.0 ($\Box$), 2.2 ($\lozenge$) and
2.8 fm$^{-1}$ ($\vartriangle$), and the NN-only part of the chiral interaction
NNLO$_{\rm sat}$ ($\triangledown$). Overall, most calculations that include 3NFs
reproduce the experimental proton radii well. For the binding energies, the
calculations with the EM(1.8/2.0) and NNLO$_{\rm sat}$ interactions reproduce
both data very well. It is important to note that without 3NFs the calculated
proton radii are about 9 -- 15$\%$ (18$\%$) smaller, while the ground states are
overbound by as much as about 24$\%$ (26$\%$) for $^{14}$C ($^{15}$C). These
results highlight the importance of comparing both experimental
observables to examine the employed interactions.

\begin{figure}[htbp]
\includegraphics[scale=0.4]{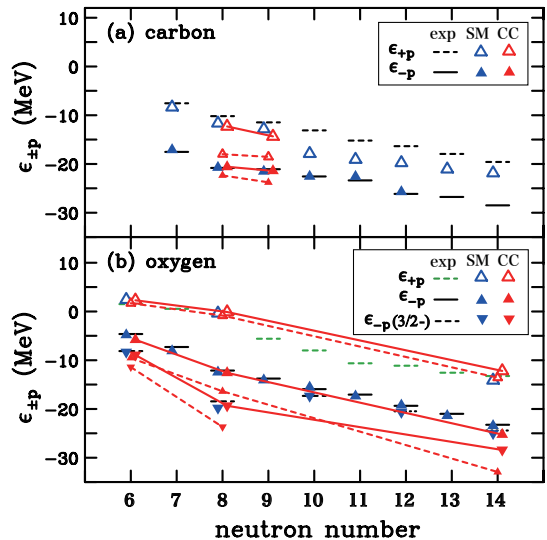}
\caption{\label{fig:be} Empirical one-proton addition ($\epsilon_{+\rm p}$) and
  removal ($\epsilon_{-\rm p}$) energies (horizontal bars) for (a) carbon, and
  (b) oxygen isotopes deduced from one-proton separation energies and the
  excitation energies of the lowest 3/2$^-$ states in the odd-even nitrogen
  isotopes. The dotted bars indicate the adopted values for the observed excited
  states in $^{19,21}$N, which have been tentatively assigned as
  $3/2^-$~\cite{SohlerPRC77}. Other experimental data are taken from
  refs.~\cite{WangCPC36,AjzenNPA523,TilleyNPA565}. The blue symbols are the
  shell model calculations using the YSOX interactions~\cite{YSOXPRC85}.
  Results of the CC calculations with and without 3NFs are shown by the
  red-solid and red-dashed lines, respectively.}
\end{figure}

\indent The importance of the Fujita-Miyazawa type~\cite{OtsukaPRL105} or the
chiral NNLO 3NFs~\cite{HagenPRL108,CipollonePRL111} in reproducing the binding
energies and the driplines of nitrogen and oxygen isotopes have been
suggested in recent theoretical studies. Here, to shed light on the role of
3NFs on the observed `subshell gap', i.e. the SO splitting in the carbon
isotopes, we investigate the evolution of one-proton separation energies
for carbon and oxygen isotopes. In Fig.~\ref{fig:be}, the horizontal bars
represent the experimental one-proton addition ($\epsilon_{+\rm p}$) and removal
($\epsilon_{-\rm p}$) energies for (a) carbon and (b) oxygen isotopes deduced
from one-proton separation energies and the excitation energies of the lowest
$3/2^-$ states in the odd-even nitrogen isotopes. The dotted bars indicate the
adopted values for the observed excited states in $^{19,21}$N, which have been
tentatively assigned as $3/2^-$~\cite{SohlerPRC77}. Other experimental data
are taken from refs.~\cite{WangCPC36,AjzenNPA523,TilleyNPA565}. For comparison,
we show the one-proton addition and removal energies (blue symbols) calculated
using the shell model with the YSOX interaction~\cite{YSOXPRC85}, which was
constructed from a monopole-based universal interaction ($V_{MU}$). Because the
phenomenological effective two-body interactions were determined by fitting
experimental data, they are expected to partially include the three-nucleon
effect and thus can reproduce relatively well the ground-state energies, drip
lines, energy levels, as well as the electric and spin properties of carbon and
oxygen isotopes. As shown in Fig.~\ref{fig:be}, the shell model calculations
reproduce the binding energies of nitrogen, oxygen, fluorine, as well as boron
and carbon isotopes around $^{14}$C very well, but underbind boron and carbon
isotopes with $N \ge 10$ by as much as 4.5 MeV.

\indent As mentioned earlier, in the absence of many-body correlations,
$\epsilon_{\pm{\rm p}}$ resemble the proton single-particle energies, and the
gap between them can be taken as the (sub)shell gap. In the following, we
consider $^{14,15}$C and the closed-shell $^{14,16,22}$O isotopes in more
detail. We computed their ground-state binding energies and those of their
neighbouring isotones $^{13,14}$B, $^{13,15,16,21}$N and $^{15,17,23}$F. We applied
the $\Lambda$-CCSD(T) and the particle-attached/removed EOM-CC methods to
compute the binding energies for the closed-(sub)shell and open-shell nuclei,
respectively. The ground-state binding energies of $^{14}$B ($2^-$) and
$^{16}$N ($2^-$) were computed using the EOM-CC method with reference to $^{14}$C
and $^{16}$O employing the charge-exchange EOM-CC
technique~\cite{EkstromPRL113}. Results of the CC calculations on $^{14,15}$C
and $^{14,16,22}$O with and without 3NFs are shown by the red-solid and
red-dashed lines, respectively. Here, we have opted for EM(1.8/2.0 fm$^{-1}$),
which yield the smallest chi-square value for the calculated and experimental
binding energies considered, as the NN+3NF interactions. For the NN-only
interaction, we show the calculations with EM(2.8 fm$^{-1}$). The calculated
$\epsilon_{-{\rm p}}$($3/2^-$) for $^{22}$O with EM(2.8 fm$^{-1}$) (and other
NN-only interactions) has an unrealistic positive value, and is thus omitted.
We found the norms of the wave functions for the one-particle (1p) $1/2^-$ and
one-hole (1h) $3/2^-$ states of $^{14}$C, and the two corresponding  1p and 1h
states of $^{15}$C ($2^-$ states in $^{14}$B and $^{16}$N) to be almost 90$\%$.
The calculations suggest that these states can be accurately interpreted by
having dominant single-particle structure, and that the gaps between these
1p-1h states resembles the `proton-subshell gaps'. It is obvious from the
figure that the calculations with the NN+3NF interactions reproduce the
experimental $\epsilon_{\pm{\rm p}}$ for $^{14,15}$C and $^{14,16,22}$O very well.
Overall, the calculations without 3NFs predict overbound `proton states', and
in the case of $^{14,15}$C, much reduced `subshell gaps'. The present results
show that $^{14}$C is a {\it double-magic} nucleus, and $^{15}$C a
{\it proton-closed shell} nucleus.

\indent Finally, we would like to point out that an inert $^{14}$C
core, built on the $N$=8 closed shell, has been postulated to explain
several experimental data for $^{15,16}$C. 
For instance, a $^{14}$C+n model was successfully applied~\cite{HassPLB59} to
explain the consistency between the measured $g$-factor and the 
single-particle-model prediction (the Schmidt value) of the excited 5/2$^+$
state in $^{15}$C. Wiedeking {\it et al.}, on the other hand, have 
explained~\cite{WiedekingPRL100} the small $B$($E2$) value in $^{16}$C assuming
a $^{14}$C+n+n model in the shell-model calculation. In terms of spectroscopy
studies using transfer reactions, the results from the
$^{14}$C(d,p)$^{15}$C~\cite{GossPRC12} and
$^{15}$C(d,p)$^{16}$C~\cite{WuosmaaPRL105} measurements are also consistent with
the picture of a stable $^{14}$C core. On the proton side, a possible
consolidation of the 1$p_{3/2}$ proton subshell closure when moving from
$^{12}$C to $^{14}$C was reported decades ago from the measurements of the
proton pick-up (d,$^3$He) reaction on $^{12,13,14}$C targets~\cite{MairleNPA253},
consistent with shell model predictions. An attempt to study the
ground-state configurations with protons outside the 1$p_{3/2}$ orbital in
$^{14,15}$C has also been reported~\cite{BedoorPRC93} very recently. To further
investigate the proton subshell closure in the neutron-rich carbon isotopes,
more experiments using one-proton transfer and/or knockout reactions induced by
radioactive boron, carbon and nitrogen beams at facilities such as ATLAS, FAIR,
FRIB, RCNP, RIBF and SPIRAL2 are anticipated.

\paragraph{Acknowledgements}
We thank T.~Shima, H.~Toki, K.~Ogata and H.~Horiuchi for discussion, and
K.~Hebeler for providing matrix elements in Jacobi coordinates for 3NFs at NNLO.
H.J.O. and I.T. acknowledge the support of A. Tohsaki and his spouse. D.T.T.
and T.T.N. appreciate the support of RCNP Visiting Young Scientist Support
Program. This work was supported in part by Nishimura and Hirose
International Scholarship Foundations, the JSPS-VAST Bilateral Joint Research
Project, Grand-in-Aid for Scientific Research Nos. 20244030, 20740163, and
23224008 from Japan Monbukagakusho, the Office of Nuclear Physics, U.S.
Department of Energy, under grants DE-FG02-96ER40963, DE-SC0008499 (NUCLEI
SciDAC collaboration), the Field Work Proposal ERKBP57 at Oak Ridge National
Laboratory (ORNL), and the Vietnam government under the Program of Development
in Physics by 2020. Computer time was provided by the Innovative and Novel
Computational Impact on Theory and Experiment (INCITE) program. This research
used resources of RCNP Accelerator Facility and the Oak Ridge Leadership
Computing Facility located at ORNL, which is supported by the Office of Science
of the Department of Energy under Contract No. DE-AC05-00OR22725.



\vspace{-4mm}
\bibliography{apssamp.bib}

\end{document}